\documentclass[english,aps,prb,showpacs,preprint]{revtex4}
\usepackage[T1]{fontenc}
\usepackage[latin9]{inputenc}
\usepackage{amsmath}
\usepackage{graphicx}
\usepackage{amssymb}

\makeatletter

\providecommand{\LyX}{L\kern-.1667em\lower.25em\hbox{Y}\kern-.125emX\@}
\providecommand{\tabularnewline}{\\}

\makeatother

\usepackage{babel}

\begin{document}

%%\preprint{08/06/2010}

\title{Enhanced Static Approximation to the Electron Self-Energy Operator for Efficient Calculation of
Quasiparticle Energies}

\author{Wei Kang and Mark S. Hybertsen}

\affiliation{Center for Functional Nanomaterials, Brookhaven National Laboratory,
Upton, NY 11973}
\begin{abstract}
An enhanced static approximation for the electron self energy operator
is proposed for efficient calculation of quasiparticle energies. 
Analysis of the static COHSEX approximation originally
proposed by Hedin shows that most of the error
derives from the short wavelength contributions
of the assumed adiabatic accumulation of the Coulomb-hole.
A wavevector dependent correction factor can be incorporated
as the basis for a new static approximation.
This factor can be approximated by a single scaling
function, determined from the homogeneous electron gas model.
The local 
field effect in real materials is captured by a simple ansatz based on symmetry
consideration. 
As inherited
from the COHSEX approximation, the new approximation presents a Hermitian self-energy 
operator and the summation over empty states is eliminated from the evaluation of the self energy operator. 
Tests were  conducted comparing the new approximation to GW calculations
for diverse materials ranging from crystals and nanotubes.
The accuracy for the minimum gap is about 10\% or better. 
Like in the COHSEX approximation, the occupied bandwidth is overestimated.
\end{abstract}
\date{\today}
\pacs{71.15.Qe 71.10.-w 73.22.-f}
\maketitle

\section{introduction}

Understanding the electronic excitation energies of a material system is
fundamental to a broad array of material properties.  Formally,
these correspond to the spectra associated with electron removal 
and electron addition.  In practice, the electronic excitations are the starting
point for understanding many phenomena, e.g. through the density of states
at the Fermi energy of a metal, the minimum energy gap and associated band
effective masses in a semiconductor or the frontier energy levels
in a nanoscale junction that control electron tunneling.  While the
electronic excitation energies can be very often interpreted within an
independent electron picture, the many-body treatment of the 
electron-electron interaction remains fundamental to the predictive calculation
of the quasiparticle energies \cite{Hedin1969}.
Density functional theory (DFT) \cite{Hohenberg1964,Kohn1965}
has been widely successful in the prediction of the ground-state derived
properties of a wide array of materials systems.  However, the corresponding
effective single particle eigenvalues that emerge from the Kohn-Sham
equations are not generally justified to be interpreted as quasiparticle energies
and in practice there are significant errors such as the substantial underestimation
of semiconductor band gaps \cite{Jones1989}.
In a many-body perturbation theory approach, the central quantity in the
theory is the non-local, energy dependent electron self energy operator.
The GW approximation for the electron self energy introduced by Hedin \cite{Hedin1965} 
has been widely exploited for predictive calculations of quasiparticle 
energies in real materials \cite{Aryasetiawan1998,Aulbur1999,Onida2002}.

The substantial extra complexity associated with calculating the non-local,
energy dependent self energy operator and then using it to solve for
quasiparticle properties inspired early efforts to find simplifying approximations, 
for example the local approximation suggested by Sham and Kohn \cite{Sham1966}
and the static COHSEX approximation of Hedin \cite{Hedin1965}.
However, since the first successful implementations for real
materials \cite{Hybertsen1985b,Hybertsen1986,Godby1987}
it has been clear that both non-locality and energy dependence
of the self energy operator play an essential role for accurate results.
Subsequently GW calculations have been employed
as a first-principles method for a broad array of real materials \cite{Aryasetiawan1998,Aulbur1999,Onida2002},  
the physical systems ranging from bulk semiconductors\cite{Zhu1991}
to nanoclusters\cite{Saito1989,Onida1995,Reining2000} and nanotubes\cite{Spataru2004,Miyake2005}. 
As the field has advanced,
the methodology has been extended to include approximate selfconsistency
in the Green's function \cite{Ku2002,Bruneval2006,Kotani2007,Shishkin2007} and the role of vertex corrections is 
currently under debate \cite{Shirley1996,Holm1998,Shishkin2007a},
both at the expense of further computational burden.

Several factors contribute to the complexity of GW-based calculations.
The self energy operator is fundamentally non-local and energy dependent.
Furthermore, the usual formulation of the calculations for both the
screening of the Coulomb interaction and then the electron self energy operator
involve a summation over empty states of the reference Hamiltonian.
In practical calculations, generation of the corresponding orbitals 
requires considerably more effort than conventional ground state calculations where 
the diagonalization can be essentially restricted to the occupied space.
Then convergence with respect to the summation over empty states must be
carefully checked for each application.
Analysis of the algorithms in use shows that the computational burden grows
as the fourth power of the system size \cite{Aulbur1999}, although if the short range
of the non-locality of some of the operators can be exploited, the
scaling improves to essentially quadratic \cite{Aulbur1999,Rieger1999}.

Recently there is a resurgence
in research directed to improving algorithms
so that the GW method can be applied to more complex systems.
Proposals have been made for simplified closures of the summation on empty
states for the polarizability and self energy operator\cite{Bruneval2008}.
Alternative, efficient basis sets to represent the operators have been explored \cite{Lu2008,Umari2009}.
Several schemes to reformulate the perturbation theory
using iterative techniques \cite{Baroni2001}
to avoid explicit calculation of the empty states have been put
forward\cite{Wilson2008,Wilson2009,Nguyen2009,Umari2010,Giustino2010}.
Although these schemes do not generally alter the scaling with system size,
they do show potential for significant changes in the prefactor.
This makes the treatment of larger systems feasible in practice.

An alternative approach to simplify the calculations follows the
route of physically motivated approximations or models.
For example, proposals have been put forward to model the dielectric
matrices for solids including local fields \cite{Ortuno1979,Hybertsen1988}.
The local approximation to the self energy operator \cite{Sham1966} has been extended
to semiconductors through models that incorporate the incomplete screening \cite{Wang1983,Gygi1989}.
The COHSEX approximation of Hedin eliminates the summation over empty
states for the self energy operator and has the added benefit of
being a static operator, 
a particular simplification for self consistent calculations \cite{Bruneval2006,Kotani2007}.
However, the magnitude of the self energy operator is too large,
raising concerns for energy level alignment at interfaces,
and  in application to semiconductors, it tends to substantially overestimate band gaps.
One proposal suggested  that
the dynamical contribution missing in the static COHSEX model could
be captured by a linear expansion of the energy dependence in the
self-energy and a model dielectric response without extra computational
costs \cite{Bechstedt1992}. 
Finally, the hybrid functional approach in DFT, in which a fixed fraction of the 
exchange operator based on the bare Coulomb interaction (or a range truncated interaction)
is explicitly included,
empirically results in improved values for the band gaps in bulk semiconductors
and insulators \cite{Muscat2001,Heyd2005}.
For the present discussion, this approach can be viewed as an approximation that
captures some of the nonlocality of the screened exchange term in the electron self energy.
However, the residual does not capture the environment dependence of the screening
and hence important physical effects such as the image potential contribution at a surface \cite{Neaton2006}.
A recently proposed semilocal effective potential approach will likely present similar problems \cite{Tran2009}.
Overall, previous approaches have been limited in accuracy 
and in applicability to diverse systems.

A static model for the electron self energy operator
offers some compelling advantages,
including the orthonormality of the quasiparticle wavefunctions,
simplification of a selfconsistent approach
and ease of application to more complex systems
such as nanoscale junctions.
This motivates us to revisit the COHSEX approximation and to
investigate the sources of error. Starting with a careful re-examination of the homogeneous
electron gases (HEG) case, we find that most of the error of
comes from the
Coulomb-hole (COH) contributions. Physically the error originates
from the assumed adiabatic accumulation of the ``Coulomb-hole'' of the dynamic
screened Coulomb interaction. 
This error is wave-length dependent: it is negligible at long
wave-length but introduces a factor of two error at short wave-length.
A similar behavior can be seen for the case of crystalline silicon.
With this insight, we suggest an empirical model that
incorporates a wavelength dependent correction factor
to account for the
average non-adiabatic effect.
Using the results from the HEG as a guide,
a simple universal form is proposed for this correction factor, including
local field effects in crystals. In this way, we
have devised a new approximation which inherits the advantage of efficiency
from the static COHSEX approximation but improves its accuracy,
as demonstrated for a diverse series of examples.
For crystals in particular, we show that the new approximation can
be combined with an established model for the dielectric screening \cite{Hybertsen1988},
completely eliminating the sum over empty states from the calculations.

The rest of the article is organized as follows. In Sect. II, the
static COHSEX approximation is analyzed.  Then in Sect. III, the new method
is derived as a natural correction resulting from the analysis. In
Sec. IV, the proposed static method is applied to various physical
systems. The new results are compared with the static COHSEX
approximation and full GW calculation. Section V provides a brief summary.

\section{Analysis of the COHSEX approximation}

The electron self energy operator in 
the GW approximation can be written in the energy domain as 
\begin{eqnarray}
\Sigma(\mathbf{r,r'};E)=\qquad\qquad\qquad\qquad\qquad\qquad\qquad\qquad\nonumber \\
\frac{i}{2\pi}\int dE'e^{-i\delta^{+}E'}G(\mathbf{r,r'};E-E')W(\mathbf{r,r'};E').\label{eq:2}
\end{eqnarray}
where the full one-particle Green's function $G$ and 
the dynamically screened Coulomb interaction $W$ enter \cite{Hedin1965}.
In most practical calculations the $G$ is replaced by one
derived from a reference, single particle Hamiltonian.
Often this is based on the Kohn-Sham states calculated
with an approximate exchange correlation functional, but it might
be derived from an approximate self consistent GW approach, e.g. where
the self energy operator is replaced by the COHSEX approximation \cite{Bruneval2006}
or by the approximate projection in the quasiparticle self consistent approach \cite{Kotani2007}.
With this approximation for $G$, then the real part of the self energy operator
can be easily rewritten in the form
\begin{eqnarray}
\Sigma(\mathbf{r,r'},E)=-{\displaystyle \sum_{n,\mathbf{k}}^{occ}\phi_{n,\mathbf{k}}(\mathbf{r})\phi_{n,\mathbf{k}}^{*}(\mathbf{r'})}W(\mathbf{r,r'};E-E_{n,\mathbf{k}})\nonumber \\
+\sum_{n,\mathbf{k}}\phi_{n,\mathbf{k}}(\mathbf{r})\phi_{n,\mathbf{k}}^{*}(\mathbf{r'})P{\displaystyle \int_{0}^{\infty}dE'\frac{B(\mathbf{r,r'};E')}{E-E_{n,\mathbf{k}}-E'}}.\label{eq:5}
\end{eqnarray}
The first term is the contribution from the poles of the Green's
function $G$, while the second term comes from the spectral
function $B$ of the screened Coulomb interaction $W$. The symbol
$P$ refers to the Cauchy principal value of the integration. The
first term is the dynamically screened-exchange (SEX) contribution
and the second term is the dynamical Coulomb-hole (COH) contribution \cite{Hedin1965}.

The static COHSEX approximation can be obtained formally by putting
$E-E_{n,k}\rightarrow0$ in Eq. (\ref{eq:5}): 
\begin{equation}
\Sigma_{SEX}^{static}(\mathbf{r,r'},E)=-{\displaystyle \sum_{n,\mathbf{k}}^{occ}\phi_{n,\mathbf{k}}(\mathbf{r})\phi_{n,\mathbf{k}}^{*}(\mathbf{r'})}W(\mathbf{r,r'};E=0),\label{eq:6}\end{equation}
and
\begin{equation}
\Sigma_{COH}^{static}(\mathbf{r,r'},E)=\frac{1}{2}\delta(\mathbf{r-r'})W_{p}(\mathbf{r,r'};E=0),\label{eq:7}
\end{equation}
where $W_{p} = W - v$ and $v$ is the bare Coulomb interaction.
Physically, when the self energy operator is evaluated for a specific
quasiparticle energy $E^{qp}$, then the static approximation assumes
that the magnitude of the energy $E^{qp}-E_{n,k}$
in Eq. (\ref{eq:5}) is much smaller than the characteristic energy
of the screening, e.g, the plasmon energy \cite{Hybertsen1986}.
Alternatively, one can write the approximate formulae in the
time domain as 
\begin{eqnarray}
\Sigma^{COHSEX}(\mathbf{r,r'};t)=\qquad\quad\qquad\qquad\qquad\qquad\qquad\nonumber \\
iG(\mathbf{r,r'};t)[v(\mathbf{r,r'};t+\delta^{+})+\delta(t)W_{p}(\mathbf{r,r'};E=0)],\label{eq:8}
\end{eqnarray}
where $W_{p}(\mathbf{r,r'};E=0)=\int_{-\infty}^{+\infty}dt[W(\mathbf{r,r'};t)-v(\mathbf{r,r'};t)$].
Noting that the $W(\mathbf{r,r'};t+\delta^{+})$
in the original GW formula can be recast as $W(\mathbf{r,r'};t+\delta^{+})=v(\mathbf{r,r'};t+\delta^{+})+[W(\mathbf{r,r'};t+\delta^{+})-v(\mathbf{r,r'};t+\delta^{+})]$,
it is clear that the only approximation made in the static COHSEX
approximation is the substitution of $[W(\mathbf{r,r'};t+\delta^{+})-v(\mathbf{r,r'};t+\delta^{+})]$
by $\delta(t)W_{p}(\mathbf{r,r'};E=0)$.
Physically this approximation replaces the time-dependent screened interaction
with an instantaneous interaction which is
the adiabatic accumulation of the ``Coulomb hole''
of the time-dependent screened Coulomb interaction \cite{Hedin1965,Hedin1969}. 

\begin{figure}
\includegraphics[scale=0.31]{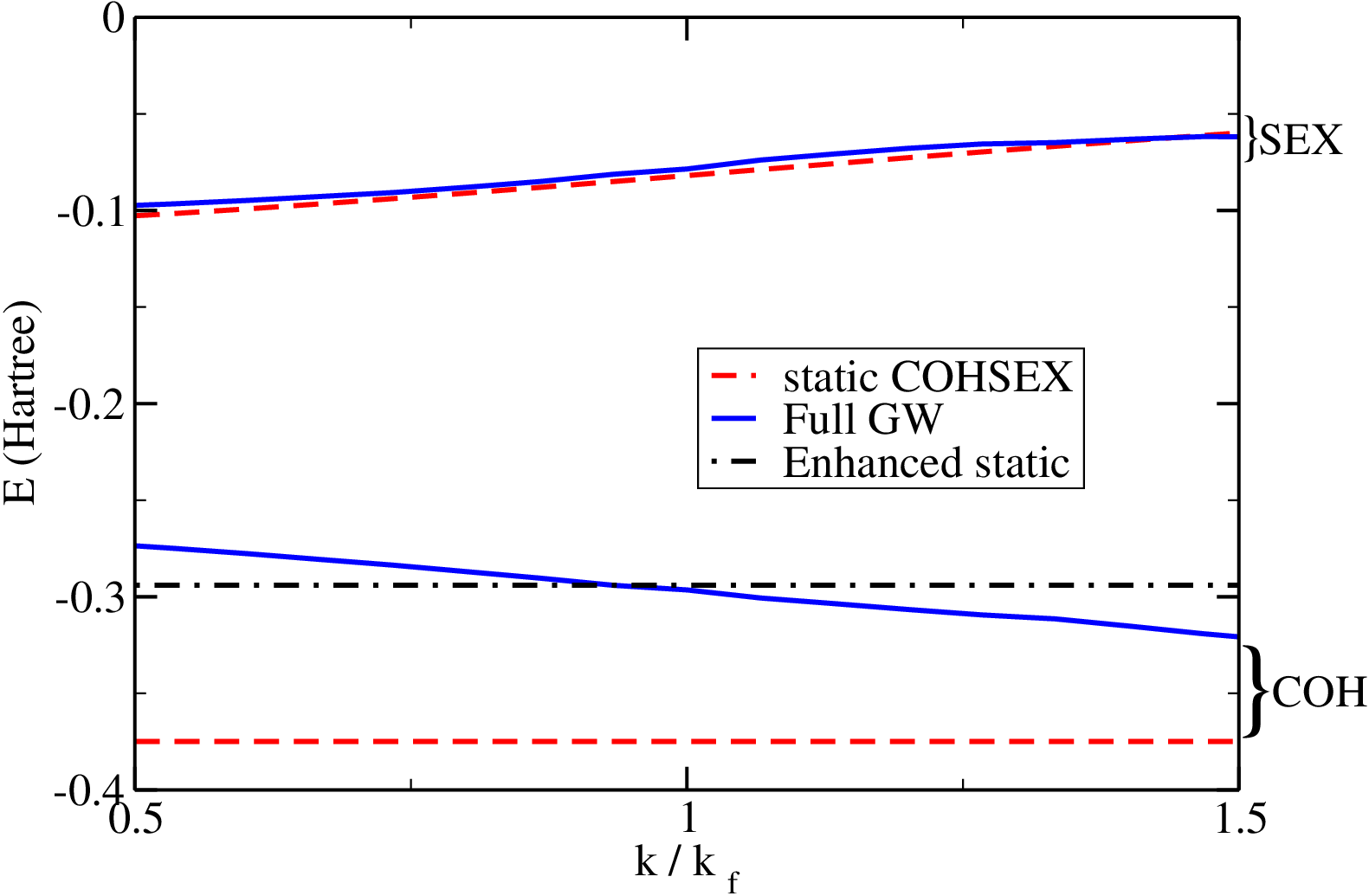}

\caption{Error analysis of the self energy $\Sigma(k,E_{k})$ for the homogeneous
electron gas with density parameter $r_{s}=2.0$. Here $k_{f}$ is the magnitude of Fermi
wave vector. The $\Sigma$ is split into COH and SEX contributions,
as indicated in the figure. Solid curves are results from full GW
calculation, and dashed curves are calculated from the static COHSEX
approximation. For references, the COH contribution from the new
enhanced static approximation is also displayed in
the figure as a dash-dot line.
 \label{fig:1}}

\end{figure}

The adiabatic accumulation of the ``Coulomb hole'' $W_{p}$ has different
influence on the SEX and COH contributions, although 
this is difficult to assess analytically.
Numerically it can be shown that
most of the error in the static COHSEX approximation comes from the
COH contribution. The SEX term in the approximation is relatively close
to the full GW calculations. 
For example, Fig. \ref{fig:1} displays the COH and SEX contributions of the self
energy $\Sigma(k,E_{k})$, evaluated with the full-frequency Lindhard
dielectric function for the homogeneous electron gas of density parameter $r_{s}=2.0$.
The SEX contribution is around -0.1 Hartree and increases slowly with
$k$. Compared with the full GW calculation, the static COHSEX approximation
slightly underestimate the SEX contribution and the difference is
less than $5.0\times10^{-3}$ Hartree (0.14 eV) for $k$ from $0.5k_{f}$
to $1.5k_{f}$. On the Fermi surface, the difference is $3.4\times10^{-3}$
Hartree (0.093 eV). The COH contribution is independent of $k$ in
the static COHSEX approximation, due to the locality in space, 
while in the full GW calculation,
it has modest dispersion. Most striking is the substantial error 
in the overall magnitude of the COH contribution ranging
from 0.10 Hartree (2.7 eV) at $k=0.5k_{f}$ to 0.052 Hartree (1.4
eV) at $k=1.5k_{f}$. On the Fermi surface the error is 0.078 Hartree
(2.1 eV). 

Similar trends are also observed in real materials.
Table \ref{tab:1} shows the SEX and COH contributions for bulk Si
and LiCl, as well as argon in the solid state, evaluated at the quasiparticle
energies of the highest occupied states
and the lowest empty states.
(For Si, the conduction band minimum is slightly lower, 
located along the $\Delta$ 
line in the Brillouin zone.)
Compared with the results of full GW calculations (described
in more detail below in Sect IVA), the static
COHSEX approximation has slight deviation for the SEX contributions (up to 0.3 eV),
while the magnitude of the COH contributions are overestimated by 1 to 3 eV.
For example, the COH contribution to the valence band
maximum (VBM) of LiCl, is wrong by 2.6 eV.

\begin{figure}
\includegraphics[scale=0.33]{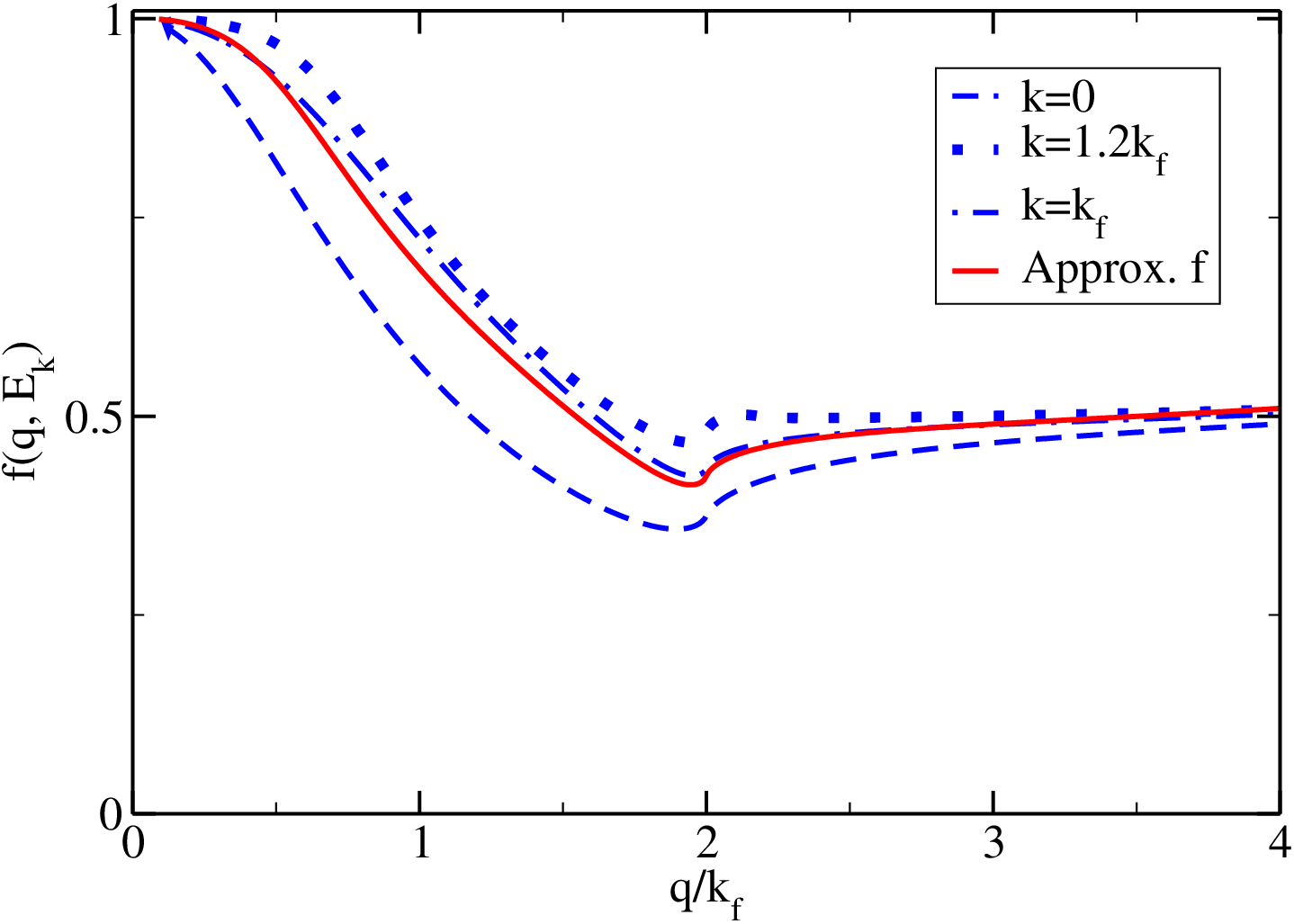}

\caption{Typical distributions of $f(q,E_{k})=\frac{W_{p}^{full}(q,E_{k})}{W_{p}(q,E=0)}$
of homogeneous electron gas ($r_{s}$=2.0). The reciprocal of $f$
represents the deviation of the adiabatic accumulation of the ``Coulomb
hole'' $W_{p}$ as a function of wave vector $\mathbf{q}$ in the
static COHSEX approximation. The dotted curve in the figure is the
distribution for the state $k=1.2k_{f}$, the dash-dot curve is for
$k=k_{f}$, and the dashed curve is for the state $k=0$. Also displayed
in the figure (displayed as a solid curve) is the approximated distribution
$f^{*}$ in Eq. (\ref{eq:11}) adopted in the new enhanced approximation.
\label{fig:2}}

\end{figure}

To get more insight to the errors, we compare the matrix element
of $\Sigma_{COH}$ for the full GW calculation
\begin{equation}
\left\langle \mathbf{k}|\Sigma_{COH}^{Full}(E_{k})|\mathbf{k}\right\rangle =\int d\mathbf{q}\left[P\int_{0}^{\infty}\!\! dE'\frac{B(q,E')}{E_{k}-E_{q+k}-E'}\right],\label{eq:9}
\end{equation}
to that for the static COHSEX approximation 
\begin{equation}
\left\langle \mathbf{k}|\Sigma_{COH}^{static\, COHSEX}(E_{k})|\mathbf{k}\right\rangle =\int d\mathbf{q}\left[W_{p}(q,E=0)\right].\label{eq:10}
\end{equation}
Both equations show the wave vector decomposition of the contributions
to the COH term.
Implicitly, Eq. (\ref{eq:9}) defines the full accumulation of the ``Coulomb
hole,'' $W_{p}^{full}(q,E_{k})$,
the counterpart of $W_{p}(q,E=0)$ in Eq. (\ref{eq:10}). 
The ratio $f(q,E_{k})=\frac{W_{p}^{full}(q,E_{k})}{W_{p}(q,E=0)}$
reflects the deviation of the adiabatic accumulation at each wave
vector $\mathbf{q}$. 

Figure \ref{fig:2} displays typical distributions of $f(q,E_{k})$
for the HEG. For small wave vector $q$ (the long wavelength limit),
the ratio $f$ approaches to 1, suggesting that the adiabatic accumulation
in the static COHSEX approximation works well. But for large $q$
(the short wavelength limit) the ratio approaches 0.5 asymptotically,
which indicates that the adiabatic accumulation exceeds the $W_{p}^{full}$
by a factor of 2. This large error in the short wavelength limit traces to
the fact that screening does not follow the rapid motion of electrons
at large $q$. In between, the ratio $f$ drops smoothly from 1 at
$q=0$ to a value close to 0.5 at $q=2k_{f}$; for $q>2k_{f}$
the ratio $f$ changes slowly. 
As seen in Fig. \ref{fig:2}, this behavior depends weakly on $k$.
We have also investigated $f(q,E_{k})$ for values of density parameter $r_s = 1 - 6$.
Provided the wavevector is scaled by the Fermi wavevector, the variation in the
curves spans a similar range to the $k$ dependence already shown.
While the results in Fig. \ref{fig:2} are based on full numerical calculations in the HEG, 
the same picture can be derived quite
directly from the asymptotic behavior of the Overhauser plasmon pole model\cite{Overhauser1971}. 
It follows from
the fact that as $q\rightarrow0$ the plasmon frequency $\omega_{q}$
approaches to the classical plasma frequency $\omega_{P}$, and as
$q\rightarrow\infty$ the effective pole frequency $\omega_{q}$ goes to $q^{2}/2m$.

To probe an example of a semiconductor,
the analogous calculation is performed for bulk silicon.
In crystals, the screened Coulomb interaction
is a function of $r$ and $r'$ separately, not just the difference (as it is in the HEG).
For crystals, then $W_{p}$ and the generalization of $f$ that we require, 
are functions defined on discrete points
$\mathbf{q}+\mathbf{G}$ and $\mathbf{q}+\mathbf{G'}$, where $\mathbf{q}$
is a wave-vector in the first Brillouin zone, and $\boldsymbol{G}$
and $\boldsymbol{G'}$ are reciprocal lattice vectors.
In the generalizations of Eqs. (\ref{eq:9}) and (\ref{eq:10})
we focus on the contribution of the diagonal elements (where $\mathbf{G}=\mathbf{G'}$)
and consider the matrix elements of the
self energy operator for the valence band maximum (VBM, $\Gamma'_{25v}$) of bulk silicon.
The necessary calculations for the screened Coulomb interaction and
the GW approximations are performed as described below (Sect. IV).
The results are plotted again in the form of a correction factor
as a function of $|\mathbf{q+G}|$ in Fig. \ref{fig:3}. 
The wavevector scale is normalized by
$k_{VBM}=\sqrt{\frac{2 m}{\hbar^2}\left\langle \phi_{VBM}|\frac{-\hbar^2\mathbf{\nabla^{2}}}{2m}|\phi_{VBM}\right\rangle }$, a simple analoge of the
Fermi wavevector in the HEG.
This effective correction factor shows a similar overall behavior as in the HEG, 
but at each $|\mathbf{q+G}|$ point, $f$
can have multiple values. This reflects the orientational anisotropy
in real materials. Using $k_{VBM}$ as the scale, the shape of $f$ displayed in Fig. \ref{fig:3} closely
resembles that in the HEG. 

\section{Enhanced static approximation}

From the results in Sect. II,  a strategy to improve the accuracy emerges:
simply include a correction factor to the adiabatic $W_{p}(E=0)$
in Eq.(\ref{eq:7}). Ideally, the factor is just the wavevector resolved
and energy dependent ratio $f(q,E_k)$.
However, the results of Fig. \ref{fig:2} suggest that the density and the $k$ or $E_k$
dependence of the correction factor is not large, except for the scaling of the wavevector $q$.
Furthermore, the possibility to drop the $E_k$ dependence results in an
energy independent (static) model for the self energy operator.
Therefore a universal function $f^{*}$
is proposed. A convenient Pade form for
$f^{*}$ is chosen and fit to the $f(q,E_{f})$ for HEG of
$r_{s}$=1.0:  

\begin{widetext}
\begin{equation}
f^{*}(x)=\frac{1+1.9085x-0.542572x^{2}-2.45811x^{3}+3.08067x^{4}-1.806x^{5}+0.410031x^{6}}{1+2.01317x-1.55088x^{2}+1.58466x^{3}+0.368325x^{4}-1.68927x^{5}+0.599225x^{6}}, \label{eq:11}
\end{equation}
\end{widetext} 
where $x$ represents the dimensionless wave number $q/k_{f}$. It
is also displayed as the solid curve in Fig. \ref{fig:2}. 

For the HEG, the enhanced static approximation retains the usual static screened
interaction term and alters the Coulomb hole term:
\begin{equation}
%\begin{cases}
\Sigma_{COH}^{new}=\frac{1}{2}\delta(\mathbf{r}-\mathbf{r'})\int d\mathbf{q}e^{-i\mathbf{q\cdot r}}W_{p}(q,E=0)f^{*}(q/k_{f}).
%\end{cases}
\label{eq:12}
\end{equation}
The improvement of the new approximation is easy to verify for the HEG,
as illustrated in Fig. \ref{fig:1} for $r_s = 2.0$.
The error from the COH contribution decreases
to 0.07 eV from 2.1 eV at the Fermi surface.
Examining the range $r_{s} = 1 - 6$,
the error remains within 0.2 eV at
the Fermi surface. That range of $r_{s}$
represents typical electronic densities in most bulk materials. 
Since this static approximation to the Coulomb hole term
remains local in space, it has no dispersion, as seen in Fig.\ref{fig:1}.
The occupied bandwidth will still be overestimated in this
new static approximation.

\begin{table}
\begin{tabular}{ccccccccccc}
\hline 
 &  &  & \multicolumn{2}{c}{GW} &  & \multicolumn{2}{c}{COHSEX} &  & \multicolumn{2}{c}{New Static}\tabularnewline
 &  &  & $\Sigma_{SEX}$ & $\Sigma_{COH}$ &  & $\Sigma_{SEX}$ & $\Sigma_{COH}$ &  & $\Sigma_{SEX}$ & $\Sigma_{COH}$\tabularnewline
\hline
\hline 
Si & $\Gamma_{25v}'$ &  & -3.83 & -8.28 &  & -3.91 & -10.43 &  & -3.91 & -8.21\tabularnewline
(bulk) & $X_{1c}$ &  & -1.74 & -7.40 &  & -2.06 & -8.72 &  & -2.06 & -7.25\tabularnewline
 &  &  &  &  &  &  &  &  &  & \tabularnewline
LiCl & $\Gamma_{15v}$ &  & -8.59 & -8.24 &  & -8.69 & -10.84 &  & -8.69 & -7.91\tabularnewline
(bulk) & $\Gamma_{1c}$ &  & -1.90 & -6.34 &  & -2.25 & -7.06 &  & -2.25 & -5.76\tabularnewline
 &  &  &  &  &  &  &  &  &  & \tabularnewline
Ar & $\Gamma_{15v}$ &  & -12.77 & -7.24 &  & -12.79 & -10.07 &  & -12.79 & -6.91\tabularnewline
(bulk) & $\Gamma_{1c}$ &  & -1.24 & -4.01 &  & -1.56 & -4.25 &  & -1.56 & -3.55\tabularnewline
\hline
\end{tabular}

\caption{The SEX and COH contributions to the matrix elements of the self energy $\Sigma$ for states
that define the energy gap for bulk silicon, bulk lithium chloride,
and argon in its solid phase, all in eV. Results from the full GW (with
the GPP model), the static COHSEX approximation, and the new enhanced static
approximations are presented. The self energies in the full GW calculations are evaluated
at the corresponding quasiparticle energies.
\label{tab:1}}

\end{table}

\begin{figure}
\includegraphics[scale=0.33]{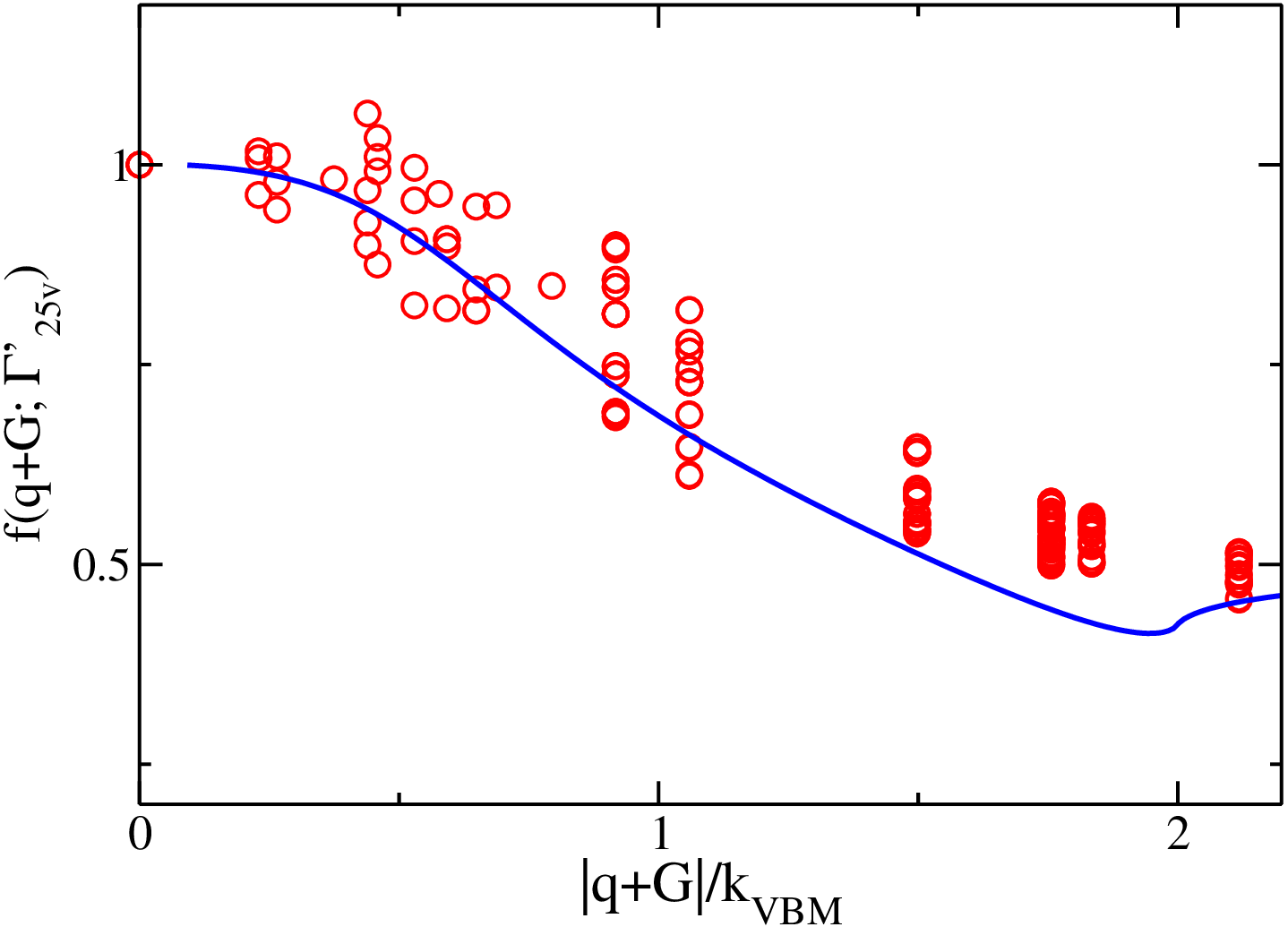}

\caption{Diagonal elements of the ratio $f(\mathbf{q}+\mathbf{G},\mathbf{q}+\mathbf{G'};\Gamma'_{25v})$
at the valence band maximum of bulk silicon as a function of $|\mathbf{q+G}|$,
displayed as circles in the figure. The solid curve is the approximate
$f^{*}$ specified in Eq. (\ref{eq:11}), and $k_{VBM}$ is the characteristic
wave number scale derived from the average speed of electrons at the
highest occupied electronic state of the system. \label{fig:3}}

\end{figure}

In order to extend this idea to real materials,
two factors must be addressed.
First a systematic scheme to derive a wavevector scale is
required.
We choose the scale
$k_{VBM}=\sqrt{\frac{2 m}{\hbar^2}\left\langle \phi_{VBM}|\frac{-\hbar^2\mathbf{\nabla^{2}}}{2m}|\phi_{VBM}\right\rangle }$
where $\phi_{VBM}$ refers to the highest
occupied electronic state in the system.
In the limit of the
HEG, $k_{VBM}$ goes back to $k_{f}$,
so it is a reasonable generalization.
With this scale factor  the approximate
$f^{*}$, the solid curve in Fig. \ref{fig:3}, is still a good description
of the diagonal terms in the numerically calculated corrections.

Second, the effect of local fields must be incorporated into the correction factor $f$.
We have tested several different generalizations that preserve symmetry
and reduce back to the simple form for the HEG limit.
We find that a simple ansatz where 
$f^{*}$ is only a function of $\sqrt{|\mathbf{q+G}||\mathbf{q+G'}|}/k_{VBM}$
works well in practice.
Accordingly the new static approximation for the COH term is revised to be
\begin{widetext}
\begin{equation} 
\Sigma_{COH}^{new}(\mathbf{r,r'})=\frac{1}{2}\delta(\mathbf{r}-\mathbf{r'})\sum_{\mathbf{q,G,G'}}e^{-i(\mathbf{q+G'})\mathbf{\cdot r'}}e^{i(\mathbf{q+G})\mathbf{\cdot r}}W_{p,\mathbf{G,G'}}(\mathbf{q},E=0)f^{*}(\mathbf{\frac{\sqrt{|\mathbf{q+G}||\mathbf{q+G'}|}}{\mathit{k}_{VBM}}}).\label{eq:13}
\end{equation} 
\end{widetext}
Note that the $f^{*}$ used here is exactly the same as the one defined
in Eq. (\ref{eq:11}). 
The $f^{*}$ in real materials should generally
be a function of both $\mathbf{q+G}$ and $\mathbf{q+G'}$ separately.
Our simple
ansatz  is isotropic and only
depends on the amplitudes of $\mathbf{q+G}$ and $\mathbf{q+G'}$.

\section{Results}

In order to test the proposed new static approximation,
calculations are performed for a diverse set of examples
including crystals, molecules, atoms and a carbon nanotube.
All calculations are performed
with geometrical parameters obtained from experiments.
All the LDA calculations are carried out in a plane-wave
basis using the Quantum Espresso package\cite{Giannozzi2009} with
norm conserving pseudopotentials
generated by the FHI99P packages\cite{Fuchs1999} using the Troullier-Martins
method\cite{Troullier1991}. The pseudopotentials are taken from the
website of ABINIT\cite{Gonze2002,Gonze2005}. In the full GW calculations,
the GPP model is used \cite{Hybertsen1986}.
Only the first order energy correction to the diagonal elements are
calculated. No further updates of spectra are included in the calculations.
The new static approximation is applied with the same statically screened
Coulomb interaction used in the full GW calculations.
For several bulk materials, the
fundamental gaps are also calculated using model dielectric matrices to
obtain the statically screened Coulomb interactions \cite{Hybertsen1988}.
This approach completely eliminates any explicit summations on empty states in the calculation.

\begin{table}
\begin{tabular}{ccccccccccc}
\hline 
 &  & Si &  & C &  & LiCl &  & GaAs &  & Ar\tabularnewline
\hline
\hline 
a$_{0}$ (nm) &  & 0.543\footnote{Ref.~\onlinecite{Kittelbook}}&  & 0.357\footnotemark[1] &  & 0.513\footnote{Ref.~\onlinecite{Madelungbook2}} &  & 0.565\footnotemark[2] &  & 0.531\footnotemark[1]\tabularnewline
\noalign{\vskip\doublerulesep}
$\epsilon_{\infty}$ &  & 12.0\footnote{Ref.~\onlinecite{Madelungbook}}&  & 5.5\footnotemark[3] &  & 2.7\footnotemark[2] &  & 10.9\footnotemark[2] &  & 1.6\footnote{Ref.~\onlinecite{Lefkowitz1967}}\tabularnewline
\hline
\noalign{\vskip\doublerulesep}
\end{tabular}
\caption{Experimental lattice constants $a_0$ and macroscopic dielectric 
constants $\epsilon_{\infty}$ of the bulk materials calculated, where
$\epsilon_{\infty}$ are parameters required by the model dielectric 
matrices \cite{Hybertsen1988}. \label{table:00}}
\end{table}

For all the bulk 
materials (including Si, C, solid Ar, GaAs, and LiCl), the LDA wavefunctions 
and eigenvalues are calculated with 80 Ry energy cutoff and the Brillouin 
zone is sampled by a 4$\times$4$\times$4 Monkhorst-Pack (MP)
mesh \cite{Monkhorst1976}. 
Their lattice constants are listed
in Table~\ref{table:00} together with  macroscopic
dielectric constants required by the model dielectric 
matrices.\cite{Hybertsen1988}  
In the GW calculations, the screened and unscreened 
Coulomb interaction are cut off at 40 Ry. 160 bands are used for the 
calculation of Green's function and the screened Coulomb interaction. 
In the calculations of atoms and molecules (including benzene, 
methene, and argon atom), the LDA wavefunctions and eigenvalues are 
calculated with 50 Ry energy cutoff in a cubic computational cell of 
1.323 nm (25.0 Bohr) for each side. 
In the GW calculations, the unscreened 
Coulomb inteaction is cut off at 10 Ry, while the screened Coulomb interaction
is cut off at 6 Ry. 700 LDA bands are used for the calculation of the Green's 
functions and screened Coulomb interaction. 
The LDA wavefunctions and eigenvalues of the single-wall carbon nanotube 
(SWCNT) (8, 0) is calculated in a trigonal computational 
cell of a = 2.381 nm (45.0 Bohr) 
and c = 0.421 nm (7.956 Bohr). The energy cutoff for the LDA wavefunction is 
60 Ry and the Brillouin zone is sampled by a 1$\times$1$\times$8 MP grid. The 
unscreened Coulomb interaction is cut off at 20 Ry. A cutoff of 6 Ry and
1000 bands are used to calculate the screened Coulomb interaction and Green's
function.
With the choice of above computational parameters, the GW energy gap are 
expected to converge within 0.1 eV for bulk materials and within 0.2 eV 
for other systems. For the SWCNT system, our computational 
parameters are slightly different from  
the previous calculation \cite{Spataru2004}. In particular, 
we do not enforce a cut off of the Coulomb  interation in the radial direction
to eliminate screening from tubes in neighboring cells.  
This reduces computational costs. 
Our results including that extra screening lead
 to a smaller quasiparticle energy gap, 
but the test of the new static method here is done with the same approximation.

\begin{table}
\begin{tabular}{ccccccccccc}
\hline 
\noalign{\vskip\doublerulesep}
Si &  & LDA &  & COHSEX &  & GW  &  & New Static &  & Expt.\tabularnewline[\doublerulesep]
\hline
\noalign{\vskip\doublerulesep}
$\Gamma_{1v}$ &  & -11.97 &  & -12.81 &  & -11.74 &  & -13.08 &  & -12.5$\pm$0.6\tabularnewline
\noalign{\vskip\doublerulesep}
$\Gamma'_{25v}$ &  & 0 &  & 0 &  & 0 &  & 0 &  & \tabularnewline
\noalign{\vskip\doublerulesep}
$\Gamma_{15c}$ &  & 2.56 &  & 3.86 &  & 3.35 &  & 3.45 &  & 3.4\tabularnewline
\noalign{\vskip\doublerulesep}
$\Gamma'_{2c}$ &  & 3.11 &  & 4.24 &  & 3.86 &  & 4.17 &  & 4.2\tabularnewline[\doublerulesep]
\noalign{\vskip\doublerulesep}
 &  &  &  &  &  &  &  &  &  & \tabularnewline[\doublerulesep]
\noalign{\vskip\doublerulesep}
$L'{}_{2v}$ &  & -9.63 &  & -10.35 &  & -9.57 &  & -10.50 &  & -9.3$\pm$0.4\tabularnewline[\doublerulesep]
\noalign{\vskip\doublerulesep}
$L{}_{1v}$ &  & -6.99 &  & -7.26 &  & -6.98 &  & -7.61 &  & -6.7$\pm$0.2\tabularnewline[\doublerulesep]
\noalign{\vskip\doublerulesep}
$L'{}_{3v}$ &  & -1.19 &  & -1.20 &  & -1.21 &  & -1.29 &  & -1.2$\pm$0.2,1.5\tabularnewline
\noalign{\vskip\doublerulesep}
$L_{1c}$ &  & 1.42 &  & 2.61 &  & 2.18 &  & 2.24 &  & 2.1,2.4$\pm$0.15\tabularnewline
\noalign{\vskip\doublerulesep}
$L_{3c}$ &  & 3.33 &  & 4.77 &  & 4.21 &  & 4.23 &  & 4.15$\pm$0.1\tabularnewline
\noalign{\vskip\doublerulesep}
$L'_{2c}$ &  & 7.55 &  & 9.50 &  & 8.37 &  & 8.40 &  & \tabularnewline
\noalign{\vskip\doublerulesep}
 &  &  &  &  &  &  &  &  &  & \tabularnewline
\noalign{\vskip\doublerulesep}
$X_{1v}$ &  & -7.82  &  & -8.37 &  & -7.84 &  & -8.51 &  & \tabularnewline
\noalign{\vskip\doublerulesep}
$X_{4v}$ &  & -2.85 &  & -2.86 &  & -2.86 &  & -3.10 &  & -2.9,-3.3$\pm$0.2\tabularnewline
\noalign{\vskip\doublerulesep}
$X_{1c}$ &  & 0.64 &  & 2.05 &  & 1.46 &  & 1.31 &  & 1.3\tabularnewline
\noalign{\vskip\doublerulesep}
$X_{4c}$ &  & 9.96 &  & 11.58 &  & 10.67 &  & 11.37 &  & \tabularnewline[\doublerulesep]
\hline
\noalign{\vskip\doublerulesep}
\end{tabular}

\caption{Quasiparticle energies of crystalline Si calculated with different methods. Here
``COHSEX'' refers to the static COHSEX approximation, ``GW'' refers
to full GW results with a generalized plasmon-pole model, and ``New Static''
refers to the results from the new enhanced static approximation.
Experimental results are quoted from Ref.~\onlinecite{Hybertsen1986}.
All energies in the table are presented in eV. \label{tab:2}}

\end{table}

Silicon in the diamond structure is the prototypical covalent semiconductor
crystal and a standard test case.
Since its quasiparticle wavefunctions and charge density are extended
to fill the entire volume, it is often considered as an inhomogeneous
electron gas with an energy gap in a simplified model\cite{Levine1982}.
The quasiparticle energies calculated using the new enhanced static 
approximation, the static COHSEX approximation and the full
GW calculation are summarized in Table \ref{tab:2} together with
experimental observations. 
The differences with the previous results\cite{Hybertsen1986}
in the full GW calculations come partly from higher cut-offs in the present
calculation and also because no update of the spectrum is included here. 
Relative to the valence band edge, the lowest energy conduction band states at
the $\Gamma$, $X$ and $L$ points of the Brillouin zone are 
all remarkably similar to the full GW results.
By comparison, the static COHSEX approximation places these
states 0.4-0.7 eV higher in energy than the full GW results.
The minimum band gap, involving the lowest conduction band along the $\Delta$
line at about 85\% of the distance to the $X$ point
is estimated with the new method to be 1.18 eV, slightly
smaller than the full GW result 1.32 eV, closely following the result at the $X$ point.
The results based on the new method for all the low lying empty states 
are in good agreement with experiment (about 0.1 eV or better).
Turning to the occupied states,
the new method predicts
quasiparticle energies that are systematically deeper than the full
GW calculations, by an amount that increases further from the valence band edge.
At the bottom of the valence band, the $\Gamma_{1v}$
quasiparticle energy calculated using the new method is -13.08
eV, 1.33 eV lower than the full GW results and deeper than experiment. 
This is not surprising,
since the accuracy of the new method is optimized for bands near the
Fermi energy and the trend illustrated in Fig.\ref{fig:1} for the homogeneous
electron gas also holds here. 
Referring to Table \ref{tab:1}, the final results with the new method
clearly benefit from some cancellation of errors between the SEX and COH terms.
Also, relevant for energy level alignment at interfaces,
the magnitude of the self energy at the valence band edge is -12.11 eV
in the full GW calculation, -14.34 eV for the static COHSEX and -12.12 eV for
the new model.
The error for the new model is just 0.01 as compared to 2.2 eV for COHSEX.

Lithium chloride (LiCl) is a typical ionic crystal with a rock salt
structure. Unlike silicon, the charge density and quasiparticle
wave functions of LiCl are localized around Cl$^{-}$ anions and Li$^{+}$
cations. Since the system is conceptually far from an extended electron
gas, it raises a challenge for the new method originally derived from
a HEG model.  The fundamental band
gap calculated using the new method is 8.98 eV,
very close to the full GW calculation and 1.6 eV smaller
than the static COHSEX result (Table \ref{table:3}). Like the full GW result,
the calculated value is about 0.4 eV smaller than the measured value, 
as was observed in the previous full GW calculations\cite{Hybertsen1985}.
The placement of the empty bands at the high symmetry points of
the Brillouin zone shows an accuracy, relative to the full GW calculations, similar
to the case of Si.
Also, very similar to the Si case,
the new method is systematically places the occupied states too deep.
As illustrated in Table \ref{tab:1}, the deviations for the individual SEX and COH
terms are larger. The net error in the absolute magnitude of the valence
band matrix element of the self energy operator is modest (0.3 eV),
especially as compared to the COHSEX approximation (2.7 eV).

Solid argon presents a third type of solid, with a large band gap related to the underlying energy gap
between the occupied 3p shell and the empty 4s and 4p derived bands.
The new method gives a calculated minimum band gap within 0.2 eV of the full GW result, in contrast to
the COHSEX derived gap, which is 2.3 eV larger (Table \ref{table:3}).  Other trends are similar.
In particular, there is modest cancellation between errors in the separate SEX and COH terms (Table \ref{tab:1}).
The net error in the valence band edge matrix element of the self energy operator is about 0.3 eV, as compared
to the 2.9 eV error in the COHSEX approximation.

\begin{figure}
\includegraphics[scale=0.33]{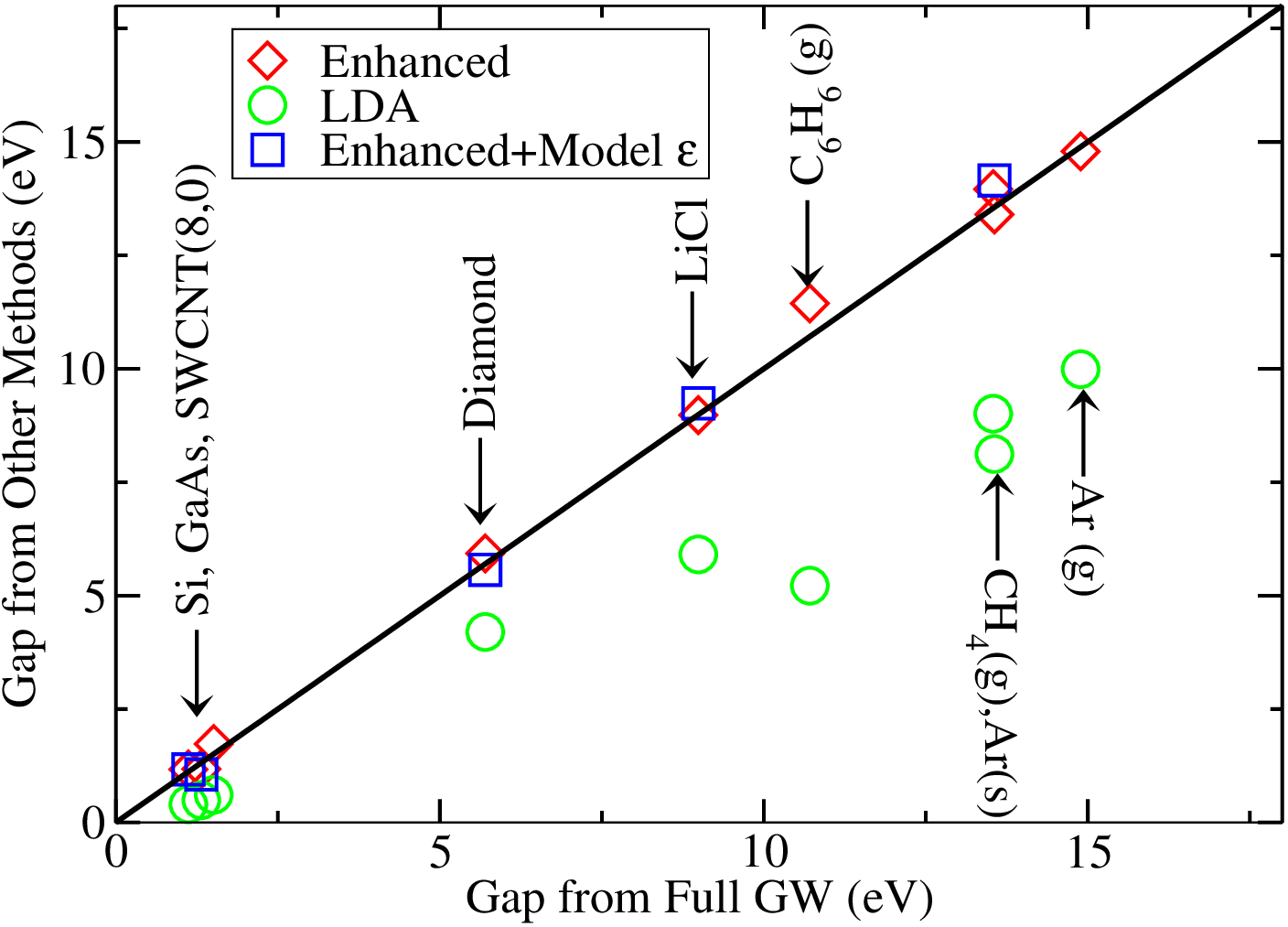}

\caption{Fundamental gaps of different physical systems compared with values
calculated from full GW method. Diamonds are gaps from the new enhanced
static approximation, squares are from a method combining the
new method and a model dielectric matrix\cite{Hybertsen1988}, and
circles are gaps from the LDA calculation. \label{fig:4}}

\end{figure}

\begin{table}
\begin{tabular}{cccccc}
\hline 
\noalign{\vskip\doublerulesep}
 & LDA & COHSEX & GW & New Static & Notes\tabularnewline[\doublerulesep]
\hline
\noalign{\vskip\doublerulesep}
Si(bulk) & 2.56 & 3.86 & 3.35 & 3.45 & $E_{g}^{\Gamma,Direct}$\tabularnewline
\noalign{\vskip\doublerulesep}
 & 0.49 & 1.91 & 1.32 & 1.18 (1.06) & $E_{g}^{Indirect}$\tabularnewline
\noalign{\vskip\doublerulesep}
 &  &  &  &  & \tabularnewline
\noalign{\vskip\doublerulesep}
C(diamond) & 5.56 & 8.35 & 7.56 & 8.00 (7.72) & $E_{g}^{\Gamma,Direct}$\tabularnewline
\noalign{\vskip\doublerulesep}
 & 4.20 & 6.99 & 5.70 & 5.93 (5.57) & $E_{g}^{Indirect}$\tabularnewline
\noalign{\vskip\doublerulesep}
 &  &  &  &  & \tabularnewline
\noalign{\vskip\doublerulesep}
LiCl(bulk) & 5.91 & 10.61 & 8.99 & 8.98 (9.24) & $E_{g}$\tabularnewline
\noalign{\vskip\doublerulesep}
 &  &  &  &  & \tabularnewline
\noalign{\vskip\doublerulesep}

GaAs(bulk) \footnote{A spin-orbital splitting correction\cite{Zhu1991} of 0.11 eV is included on the first order perturbation level.} & 0.40 & 1.43 & 1.12 & 1.17 (1.17) & $E_{g}$ \tabularnewline
\noalign{\vskip\doublerulesep}
 &  &  &  &  & \tabularnewline
\noalign{\vskip\doublerulesep}
Ar(bulk) & 8.12 & 15.85 & 13.56 & 13.40 (14.15) & $E_{g}$\tabularnewline
\noalign{\vskip\doublerulesep}
 &  &  &  &  & \tabularnewline
\noalign{\vskip\doublerulesep}
Ar (atom)  & 9.99 & 16.25 & 14.80 & 14.79 & HO/LU \tabularnewline
\noalign{\vskip\doublerulesep}
 &  &  &  &  & \tabularnewline
\noalign{\vskip\doublerulesep}
Benzene & 5.22 & 11.50 & 10.71 & 11.43 & HO/LU\tabularnewline
\noalign{\vskip\doublerulesep}
 &  &  &  &  & \tabularnewline
\noalign{\vskip\doublerulesep}
Methane & 9.01 & 15.21 & 13.54 & 13.96 & HO/LU\tabularnewline
\noalign{\vskip\doublerulesep}
 &  &  &  &  & \tabularnewline
\noalign{\vskip\doublerulesep}
SWCNT80 & 0.61 & 1.78 & 1.51 & 1.73 & $E_{g}$\tabularnewline
\hline
\noalign{\vskip\doublerulesep}
\end{tabular}

\caption{Energy gap of a variety of materials calculated using
the different methods as described at the beginning of Sec. IV. All the energies in the table are presented in eV.
Values appearing in parenthesis in the "New Static" column are calculated
with a model dielectric matrix as described in the text.
The character of the gap is noted in the final column where the
notation "HO/LU" refers to the gap between
the highest occupied molecular orbital and the lowest unoccupied molecular orbital. \label{table:3}}
\end{table}

In Fig. \ref{fig:4} and Table \ref{table:3} the fundamental gaps calculated using the new
static method are compared with full GW calculations for different types
of materials including Si (diamond structure), C (diamond), LiCl crystal,
GaAs crystal, Ar (solid), Ar (atom), benzene (molecule), methane (molecule),
and single wall carbon nanotube SWCNT(8, 0). Both calculations are
based on the same LDA wavefunctions and eigenstates. The choices of
materials in the figure cover fundamental gaps from around 1 eV up
to 15 eV, and they represent atoms, molecules, nanostructures, and
various bulk materials. As indicated in the figure, all the diamond
points (which represent results from the new method) tightly follow
the diagonal line, showing very good accuracy, about 10\% or better. 
The largest relative errors are seen for bulk Si and the SWCNT cases.
For several bulk materials
(Si, C, LiCl, GaAs, and solid Ar), we also show the results from
the combination of our new method and a model dielectric matrix\cite{Hybertsen1988}
aimed to further speed up the calculation. The results are displayed
as squares in the figure, showing that the accuracy is still maintained. 

Two examples specifically probe a pi-electron gap,
the case of the gas-phase benzene and the (8, 0) SWCNT.
In this case, the new static method and the COHSEX method
give essentially the same results.
In turn, the difference between the COHSEX results and those
from the the full GW calculation in these cases
is much more modest than for the other cases: about 0.7 eV for benzene and
0.2 eV for the SWCNT.
A closer examination of the full GW calculations
show that in this instance the contribution of the COH term to the
band gap is quite small.
This is quite different from the situation for other systems considered here.
For example, in the methane molecule, the COHSEX approximation gives
a gap that is too large by about 1.7 eV, while the new method
provides a gap that is within 0.4 eV.

\section{Summary}

A static approximation to the electron self energy offers
several technical advantages, not least of which is to maintain a hermitian
operator in the calculation of quasiparticle energies.
In addition, it offers the potential to avoid the computational burden
of converging the sum over empty states that dominates
the full application of many-body perturbation theory.
Here we have analyzed the original static COHSEX approximation
proposed by Hedin, showing that most of the errors trace to the
assumption of an adiabatic accumulation of the Coulomb hole in the
short wavelength limit.
This has lead us to propose a simple generalization
in which a single function of the scaled internal momentum in the Coulomb
hole term is used to correct this error.
Although it requires an additional ansatz to represent the local fields,
this simple, enhanced static approximation goes a surprisingly long way
to correct the errors of the original COHSEX approximation 
for application to diverse real materials, ranging from crystals
and nanotubes to molecules and atoms.
The accuracy of the new approximation may be sufficient
for a number of applications to larger scale systems.
It may also provide an efficient approximate approach for self consistent calculations. 

\begin{acknowledgments}
Work performed under the auspices of the U.S. Department of Energy
under Contract No. DEAC02- 98CH1-886. This research utilized resources
at the New York Center for Computational Sciences at Stony Brook University/Brookhaven
National Laboratory which is supported by the U.S. Department of Energy
under Contract No. DE-AC02-98CH10886 and by the State of New York. 
\end{acknowledgments}
%\bibliographystyle{apsrev}
%\addcontentsline{toc}{section}{\refname}
\bibliography{enhanced_lit}

\end{document}